\documentclass[aps,prc,showpacs,nofootinbib]{revtex4-1}
\usepackage{dcolumn}
\usepackage{graphicx}
\begin{document}

\title{$^{176}$Lu/$^{175}$Lu thermometry for Oklo natural reactors: a new look at old data}

\author{C. R.  Gould}
\affiliation{Physics Department, North Carolina State University, 
Raleigh, NC 27695-8202, USA}
\affiliation{Triangle Universities Nuclear Laboratory, Durham, 
NC 27708-0308, USA}
\email{chris_gould@ncsu.edu} 
\author{E. I.  Sharapov}
\affiliation{Joint Institute for Nuclear Research, 141980 Dubna, Moscow region,
Russia}

\date{December 12 2012}

\begin{abstract}
\begin{description}

\item[Background] Lutetium thermometry has been used to analyze Oklo 
natural nuclear reactor zones but leads to widely varying and puzzling predictions for the temperatures $T_O$ which in turn impacts bounds on time variation of the fine structure constant $\alpha$. 
\item[Purpose] We revisit results for reactor zone RZ10 in light of new measurements of the isomer branching ratio $B^g$ in $^{175}$Lu neutron capture at 5 and 25 keV. 
\item[Method] We recalculate predictions for $T_O$ as a function of $B^g$ using realistic models of the Oklo neutron flux. 
\item[Results] We find $T_O = 100 \pm 30$  C using a new value of $B^g$, in contrast to $350 < T_O < 500 $ C using the evaluated value at thermal energy.
\item[Conclusions] Lutetium thermometry can be applicable to analyses of Oklo reactor data, but a better measurement of $B^g$ with thermal neutrons is needed to confirm the reliability of temperature predictions.

\end{description}
\end{abstract}

\vspace{1pc}
\maketitle

\section {Problem of temperature in the Oklo reactors}.

\noindent The Oklo natural reactors  (see e.g. Naudet \cite{Nau91}) have proven 
to be one of the more sensitive terrestrial testing grounds for studying time variation of the 
fine structure constant $\alpha$  \cite{Shl76, Fuj00,Dam96,Chris06,Petr06,Oneg10} or time variation of the dimensionless quark mass parameter $\Xi_{q}$ of QCD \cite{Flam09}. The position of the first neutron resonance in $^{149}$Sm strongly influences  the effective cross section $\hat{\sigma}$ for neutron capture leading to burnup of $^{149}$Sm. 
The present day resonance energy  E$_0$=97.3-meV  is well known,  
but in the past may have been different,  leading to a change  (through $\hat{\sigma}$) in the isotopic abundances of various Sm isotopes in Oklo reactor wastes. From a modeling of 
the reactor parameters, and in particular from knowing the reactor temperature,  thermal and epithermal neutron spectra can be derived which can in turn be used to predict whether  the $^{149}$Sm  cross section has changed since the reactors stopped operating. While the majority of Oklo analyses have been consistent with no change, a positive effect continues to be argued for from  astronomical observations \cite{Webb11}.

Oklo analyses all reveal that the bounds on variation of $\alpha$ depend significantly on the assumed reactor operating temperature $T_O$.   
In Refs.  \cite{Shl76, Fuj00,Dam96} Maxwellian thermal neutron fluxes were used,
while in Refs. \cite{Chris06,Petr06,Oneg10}   several realistic models of the reactors zones 
were elaborated with the help of modern neutron transport codes.  
For zone RZ2, with indirect arguments,  Damour and Dyson \cite{Dam96} 
allowed for a broad interval of $T_O$ from 450 to 1000 C.
Y. Fujii et al. \cite{Fuj00} took the 
interval from 180 to 400 C for zones RZ3, RZ10,  while Gould et al. \cite{Chris06}   
preferred lower values from 200 to 300 C for zone RZ10.
Petrov et al. \cite{Petr06} argued for the value $T_O = 452\pm 55$ C 
as the temperature value 
at which their model of the   active core RZ2 became critical \footnote 
{We note  that the $T_O$  parameter depends strongly on the  $H/U$ atomic ratio, 
which in their active core model 
had rather low value, $H/U$=5.9,  as compared with $H/U$=15.6 in a subsequent model  of Onegin 
\cite{Oneg10}.
}.

Three studies have attempted to bound temperatures more directly using $^{176}$Lu/$^{175}$Lu thermometry.  
Holliger and Devillers  \cite{Holli81}, gave temperatures $T_O = $ 260 and 280  C for RZ2 and RZ3, and recently Onegin \cite{Oneg10}  
found $T_O = 182 \pm 80 $ C  for RZ3.  However, for RZ10, one of the most well characterized zones,  Hidaka and 
Holliger\cite{Hida98}  succeeded  in getting a result for only one sample $T_O =380 $ C 
while for the other three samples only a (surprisingly high) lower bound, $T_O > 1000$ C, was obtained. 
Obviously, the temperature 
$T_O$ of  Oklo   reactors remains  a very  uncertain parameter and the results of 
Ref.\cite{Hida98} are especially 
puzzling.  In what follows we will discuss some possible ways to improve the situation.   \\

In Oklo studies, the effective capture cross section  is introduced as
\begin{equation}
\hat{\sigma} = \int_0^{\infty}\,n(v)\,\sigma (v)\,v\,dv/nv_0.
\label{eq:sighat}
\end{equation}
Here $n = \int_0^{\infty} \,n(v)dv$ is the total  neutron density, and $v_0$=2200 m/sec is the velocity of a neutron at 
thermal energy $0.0253$ eV.
As an  integrated quantity,  $\hat{\sigma}$ is not  dependent on neutron velocity $v$ 
(or neutron energy $E$) but if some nuclide, like $^{176}$Lu,  has 
a neutron resonance close to thermal energy,   $\hat{\sigma}(T_{O})$  may depend on 
temperature $T_O $ through a possible temperature dependence of  the neutron density, $n(E,T_{O})$.
Examples of calculated curves $n(E, T_{O})$ for zones RZ2 and RZ10 are given in Ref.\cite{Chris06}. 
 Also
an effective neutron flux density is introduced as  $\hat{\Phi} = \int_0^{\infty} \,n(v)\,v_0\,dv$. This flux is 
different from the  integrated 'true' flux $\Phi = \int_0^{\infty} \,n(v)\,v\,dv$,  but the reaction rate $R = \hat{\sigma}\,\hat{\Phi}= \bar{\sigma}\,\Phi$ is the same since the average cross section $\bar{\sigma}= \int_0^{\infty}\,n(v)\,\sigma (v)\,v\,dv/\int_0^{\infty} \,n(v)\,v\,dv$.

We will follow the description of the Lutetium thermometry given by  
Holliger and Devillers  \cite{Holli81}. The first suggestion to use Lutetium as a sensitive indicator of the temperature of 
neutron spectra in reactors was made apparently by Westcott \cite{West58}.\\
\vspace{5mm}


\section{$^{176}$Lu/$^{175}$Lu thermometry }
\noindent  The rare earth element Lutetium has one stable isotope, $^{175}$Lu,  with natural abundance 97.401 \%, and 
and a second very long-lived isotope $^{176}$Lu  with half-life 
t$_{1/2}=37.6$ Gyr  and present day natural abundance 2.599 \% \cite{NNDC}.  Neither isotope is produced in the fission of Uranium or Plutonium.

$^{176}$Lu  has an exceptionally large thermal capture cross section, $\sigma_6= 2090 \pm 70$ b \cite{Mug06} due to a strong resonance at 141 meV. This leads to a strong temperature dependence in the rate at which $^{176}$Lu burns up. Following neutron capture, $^{176}$Lu  transforms into the stable nuclide $^{177}$Hf  after the $\beta$ decay of the product $^{177}$Lu . 
Neutron capture by $^{175}$Lu has two branches with much smaller cross sections.  
The dominant branch with $\sigma_{5}^{m} = 16.7\pm 0.4 $ b \cite{Mug06}  leads to an isomeric state
$^{176m}$Lu which decays to $^{176}$Hf with half-life of 3.6 hr. 
The minor branch with $\sigma_{5}^{g} = 6.6\pm 1.3$ b \cite{Mug06}  leads directly to the  ground state 
of $^{176}$Lu. While the large cross section for neutron capture in $^{176}$Lu serves to deplete the small fraction of $^{176}$Lu, 
the minor capture branch on the much larger fraction of $^{175}$Lu works 
to restore $^{176}$Lu. Defining $\sigma_5=\sigma_{5}^{g}+\sigma_{5}^{m}$, 
the important parameter determining the balance between depletion and restitution of $^{176}$Lu is then $B^g=\sigma_{5}^{g}/\sigma_5$, the branching ratio parameter. \footnote {Another parameter, the isomeric ratio, $IR$ is also used in the literature: $ IR = 1 - B^g$.}
According to these evaluated cross sections 
$B^g=0.28\pm 0.05$ at thermal neutron energies.

\noindent  Introducing the atomic number densites $N_i$ and 
the reaction rates $\lambda_i = \hat\sigma_i\hat{\Phi}$ (the latter  
play roles analogous to the roles of decay constants in the radioactive decay) 
and neglecting for now the $\beta$ decay of the 'stable' $^{176g}$Lu during the relatively much shorter time of Oklo 
reactor criticality,  
we write the coupled  differential equations for the time evolution of the number densities 
$N_i(t)$ of our two isotopes of interest:

\begin{eqnarray}
{dN_6\over dt}&=&-\lambda_{6}N_6+B^{g}\lambda_{5}N_5\\
{dN_5\over dt}&=&-\lambda_{5} N_5 
\end{eqnarray}
with initial conditions $ N_6(0)=N_6^{0}\exp({Dln2\over t_{1/2}})$   and $ N_5(0)=N_5^{0}$. 
In these equations, the subscripts 5 and 6 refer to $^{175}$Lu 
and $^{176}$Lu, respectively,  $N_6^{0}$ and   $N_5^{0}$  are the present day natural abundances of lutetium 
isotopes and $D=2$ Gyr  is the date of the Oklo phenomenon.  \footnote
{
The age of the Oklo natural reactors  is debated in the literature: Fujii et al \cite{Fuj00} and Dyson and Damour \cite{Dam96} use 2 Gyr. Naudet cites 1.95 Gyr \cite{Nau91} and 1.8 Gyr is used in Ref. \cite{Petr06}.
}

This system of equations has an analytical solution which we write for the time duration $t_{1}$  
of reactor criticality:
\begin{eqnarray}
N_6(t_1)=N_6^{0}\exp({Dln2\over t_{1/2}})\exp(-\lambda_{6}t_{1}) + N_5^{0}B^{g}
{\lambda_5\over \lambda_6 - \lambda_5}[\exp(-\lambda_{5}t_1) - \exp(-\lambda_{6}t_1)]\\
N_5(t_1)=N_5^{0}\exp(-\lambda_{5}t_{1}).
\end{eqnarray} 
The two terms in the right hand side of Eq.(4) represent the burnup of the initial $^{176}$Lu  with cross 
section $\sigma_6$  and its 
partial restitution after burnup of $^{175}$Lu with the partial cross section $\sigma_5 ^{g}$.
Taking the ratio and accounting for the $\beta$ decay of $^{176}$Lu after shut down of the reactor, 
we obtain the present day ratio of the lutetium isotopes in Oklo wastes as:
\begin{equation}
{N_6\over N_5}({\rm now})= {N_6^{0}\over N_5^{0}}\exp(-(\hat\sigma_6 - \hat\sigma_5) \hat{\Phi} t_1)
+ B^{g}{\hat\sigma_5\over \hat\sigma_6 - \hat\sigma_5}[1 - \exp(-(\hat\sigma_6 - \hat\sigma_5) \hat{\Phi} t_1) ]
\exp(- {Dln2\over t_{1/2}}),
\end{equation}
which depends on the temperature $T_O$ through the temperature dependence of $\hat\sigma_6$.

The neutron fluence $\hat{\Phi} t_1$ is a well characterized parameter in Oklo studies. The effective cross sections 
$\sigma_5$ and $\sigma_6$ are calculable using known resonance cross sections of Ref.\cite{Mug06}  
together with  neutron densities within realistic reactor models.  The calculation therefore leads to a clear prediction of the present isotopic ratio if the value of the branching ratio parameter $B^g$ is known.

The determination of $B^g$ has been the subject of considerable experimental effort in connection with astrophysical studies of the s-process in  stars.  The difficulty in making a precise determination  arises from the fact that its value is obtained by taking the difference between two cross sections, $\sigma_5$ and $\sigma_5^{m}$, of similar magnitude. In a 1988 summary of measurements \cite{DeLa88},  a value anywhere in the range 0.16 to 0.38 seemed possible.
To address the problem, two precision measurements of the $^{175}$Lu total capture and partial activation cross sections at 5 and 25 keV  were recently carried out \cite{Hei08,Wiss06}. The conclusions of these studies were that \cite{Hei08} 
$B^{g}= 0.117\pm 0.046$  at 5 keV and   
$B^{g}= 0.143\pm 0.020$ at 25 keV. 
These values are about a factor of two less than the value for thermal neutrons, 0.28. Such a big difference looks rather puzzling even taking into consideration a partial contribution of $p$-wave neutrons, and raises the question of whether the thermal value is actually correct.   There is some experimental evidence for a smaller value. A spectroscopic study \cite{Klay91} of thermal neutron capture found $B^g = 0.13\pm 0.03$. Also, the only neutron transmission 
measurement for which there are data with an enriched $^{175}$Lu sample, (Baston et al. \cite{Bast60}), hints at a smaller value in the following way. The thermal capture cross section they determined, $\sigma_5= 23 \pm 3$ b, was  based on subtracting a potential scattering contribution of 5.5 b from their measured total cross section. However, the  potential scattering cross section is known today to be much  larger, $\sigma_n = 7.2 \pm 0.4$ b \cite{Sears92}. Using this new value leads to $\sigma_5 = 21.3$ b, and correspondingly a smaller $B^g = 0.21 \pm 0.11$ b.

While this latter result has very large uncertainty, we believe there is  
reason to think the value of $B^g$ for thermal neutrons is not that well known, and merits further study. In what follows we evaluate Oklo reactor temperature predictions for a range of 
values from 0.10 to 0.30, and then focus on the most precise value measured to date, the 25 keV value cited above.  \\

\section{$T_O$ calculation for zone  RZ10}

\noindent To solve equation 6, we first evaluate the effective capture cross sections
$\hat{\sigma}$ for $^{175}${Lu} and $\hat{\sigma}$ for $^{176}${Lu} using the T = 20 to 500 C neutron spectra derived from MCNP calculations in our earlier work \cite{Chris06}. 
The procedure was identical to that outlined there except here the Lu-resonance shift due to a change 
in $\alpha$ is taken to be zero. For $^{175}${Lu} we include all resonances up to 49.4 eV along
with two sub-threshold resonances and the strong resonance at 96.69 eV.  For $^{176}${Lu}  
we include all resonances up to 52.13 eV. In practice, the contribution of the $E_0 =143$-meV 
resonance dominates for $^{176}$Lu, all other resonances contribute only a few percent.

\begin{table}[hb]

\caption{(Lutetium cross sections $\hat\sigma_{5}$ and $\hat\sigma_{6}$ for the Oklo RZ10 reactor at temperatures $T_O$ from 0 C to 600 C (see text). }

\label{tab: table I}

\begin{tabular}{||c|c|c||}\hline \hline
$T_O$ (C) &$\hat\sigma_{5}$ (kb) &$\hat\sigma_{6}$ (kb)\\ 
\hline

0 &  0.115 & 4.216 \\
20 & 0.115 & 4.487 \\
100 & 0.115 & 5.359 \\
200 & 0.115 & 6.310 \\
300 & 0.114 & 7.013 \\
400 & 0.114 & 7.544 \\
500 & 0.114 & 7.715 \\
600 & 0.114 & 7.750 \\

\hline
\end{tabular}
\end{table}

\begin{table}[hb]
\caption[1]{RZ10 borehole SF84 data on Lutetium isotopic abundances \cite {Hida98} and neutron fluences from \cite {Chris06}. For RZ10 $t_1 = 850$ kyr.  The metasample data point is a weighted average of the data for the four individual samples.}
\label{tab: table II}
\begin{tabular}{||c|c|c|c|c|c||} \hline\hline
Sample & Elemental Lu (ppm)&${N_{5}(now)}$& ${N_{6}(now)}$& ${N_{6}(now)\over N_{5}(now)}$ &$\hat\phi t_1\, {\rm (kb)^{-1}}$  \\ 
\hline
natural& -    & 97.401 & 2.599   &0.02668 &   -    \\
1469   &0.934 & 99.695 & 0.303   &0.00304 &0.475 \\ 
1480   &0.876 & 99.479 & 0.521   &0.00524 &0.915  \\
1485   &1.190 & 99.866 & 0.134   &0.00134 &0.645  \\
1492   &1.710 & 99.377 & 0.623   &0.00627 &0.585  \\
\hline
meta   & -    &  -     & -       &0.00418 & 0.650  \\
\hline
\end{tabular}
\end{table}

The resulting cross sections are shown in Table I. Here we also include values for 
$T_O =0$ C and $T_O =600$ C, derived from a power series extrapolation of the 20 C to 500 C results.

The Lu isotopic abundance data for reactor zone RZ10 are given by Hidaka and Holliger 
(HH) \cite{Hida98} and are shown in Table II for four different borehole depths. 
We see $^{176}${Lu} is strongly depleted in all samples although the ratio 
$ {N_6/ N_5}({\rm now})$ varies substantially, indicating, as noted by HH, heterogeneous 
operation of the reactor zone, or differences in isotope retention following reactor shut down. 
Lacking detailed further information on this situation, it appears more useful to work with 
a meta sample averaged over the four samples that are available for analysis. 
This approach was shown to be successful 
for Sm data \cite{Chris06}. Weighting each sample by its Lu elemental concentration, 
we accordingly find the meta sample ratio  $  {N_6/ N_5}({\rm now})  = 0.00418 $.

\begin{figure}[!ht]
\includegraphics[width=5in,angle=0]{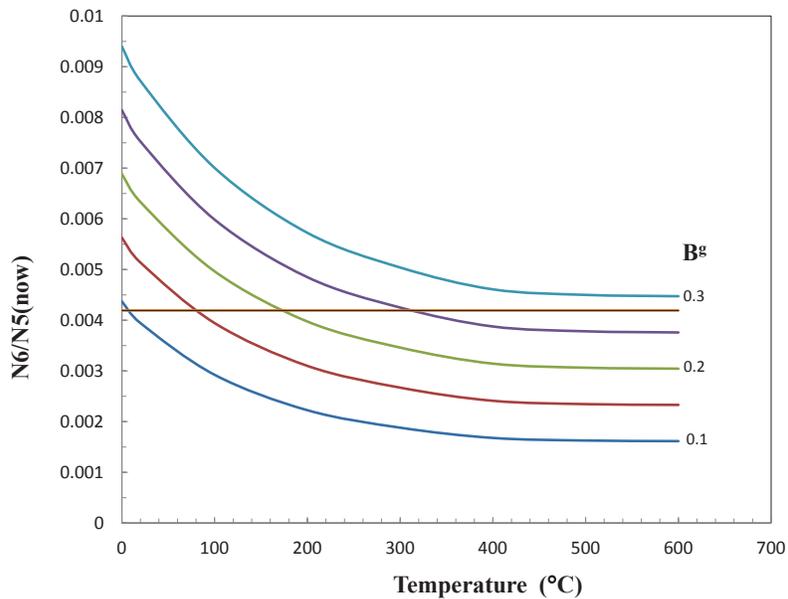}
\caption{(Color online.) Plot of solutions of Eq. 6 as a function of temperature $T_{O}$
  for $B^g$ values 0.1, 0.15, 0.2, 0.25 and 0.3, and temperatures 0 to 600 C. The meta sample value is the horizontal line $  {N_6/ N_5}({\rm now})  = 0.00419 $.}
  
\label{fig:Lu1}
\end{figure}

\begin{figure}[!ht]
\includegraphics[width=5in,angle=0]{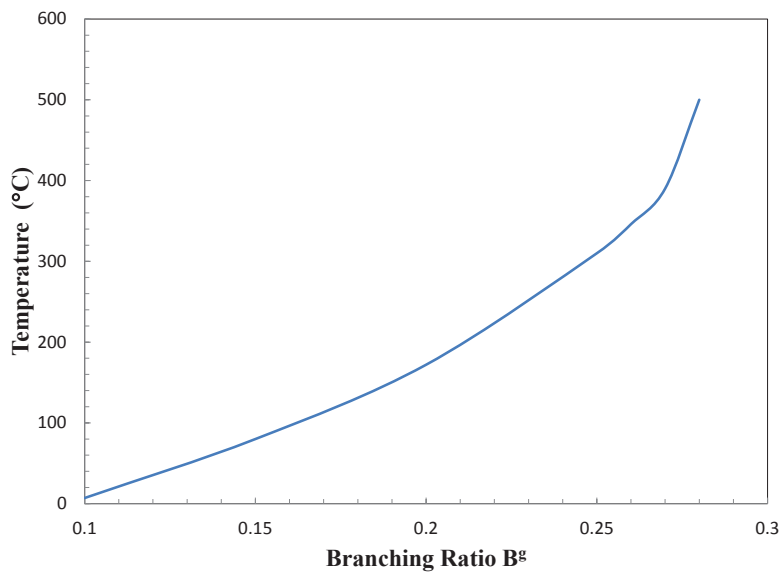}
\caption{(Color online.) Plot of the intersection temperatures for the meta sample as a function of the isomer branching ratio $B^g$. Note there is no solution above 500 C.}  
\label{fig:Lu2}
\end{figure}

This meta sample value can now be compared to the result of evaluating equation 6 for 
a range of $B^g$ values and reactor temperatures. A plot of this comparison is shown in 
Fig. 1 for $B^g$ values 0.1, 0.15, 0.2, 0.25 and 0.3, and temperatures 0 to 600 C.  Of interest 
is that for a given $B^g$ only a small range of isotope ratio values actually lead to any solution at all. Below about 350 C, the temperature is well constrained by the isotopic ratio data.  
At higher temperatures, however, the curves flatten out, leading to much reduced sensitivity, and above 500 C there is actually no solution. 

The intersection of the meta sample isotope ratio value with each curve yields a temperature 
prediction for the RZ10 reactor. A plot of these intersection values is given in Fig. 2 from 
which a temperature $T_O$ for RZ10 can be read off if $B^g$ is known. 

As noted earlier, the evaluated thermal neutron energy value \cite{Mug06} 
is $B^g=0.28\pm 0.05$ which (barely) yields a solution $350 < T_O < 500 $ C. 
The most accurately determined $B^g$ ($E_n =25$ keV) value of   $0.157 \pm 0.023$ from the study \cite{Wiss06} gives a quite different value $T_O = 100 \pm 30$  C. 

\section{Conclusions}

In this work we have identified the branching ratio parameter $B^g$ as a possible source of problems in extracting reliable temperature  bounds from Lutetium thermometry for the Oklo natural nuclear reactors. Our preference is for a lower value, but clearly there is impetus for additional study to constrain $T_O$ better.

A first possibility is to carry out an improved measurement of $B^g$ at thermal energies.  A ten percent measurement of $B^g$ is desirable and will certainly require carrying out a high precision total cross section measurement. An accuracy of 1\% has been achieved at RPI \cite{Dan11} for other rare earth element total cross sections, and similar accuracy should be feasible with an enriched $^{175}$Lu sample. The $^{176}$Lu  isomer activation cross section is less critical but should also be improved bearing in mind that the differences in the existing experimental values considerably exceed the reported uncertainties. 

There remains of course the issue of whether the meta sample, even though averaged over the samples of the RZ10 active core, still deviates 
from the true value 2 Gyr ago. Post Oklo migrations of elements in and out of the core have been discussed in the literature \cite{Gau96}. The apparent strong retention of other rare earth samples in the active cores would seem to argue against selective migration of Lutetium alone, but additional samples and data would clearly be valuable in clarifying the migration issue.

Finally there is the possibility that post processing of the Lutetium 
isotopes has occurred due to other nuclear reaction channels. This has been a topic of on-going interest in astrophysics, where an excess of $^{176}$Hf in older meteorites had been a long standing puzzle. See Thrane et al.\ \cite{Thrane} for a summary of the current situation.  It is now generally agreed that the excess is not due to an incorrect value for the $^{176}$Lu lifetime, but is instead due to excitation of the $^{176m}$Lu isomer by  gamma radiation. The source of the gamma ray flux is debated, but as discussed in Ref. \cite{Thrane}, a cosmic ray spray from a nearby supernova about 5 Gyr ago is a plausible mechanism. The fluorescence cross section for $^{176}$Lu isomer production has been intensively studied by  a number of groups (see Mohr et al.\ \cite{Mohr09}). Cross section results in the literature vary by many orders of magnitude depending on the energy of the gamma radiation. 
Although the gamma-ray fluxes in nuclear reactors are extremely low as compared to those in supernovae, additional measurements of the $^{176}$Lu photoactivation cross sections and calculations of gamma-ray fluxes in Oklo reactors are still desirable for a better understanding of the Oklo phenomenon.

\acknowledgments

We thank Art Champagne, Yuri Danon, and Alejandro Sonzogni for valuable conversations.
This work was supported by the US Department of Energy,
Office of Nuclear Physics, under Grant No.
DE-FG02-97ER41041 (NC State University).


\begin{thebibliography}{99}



\bibitem{Nau91} R. Naudet, {\em Oklo: des R\`{e}acteurs Nucl\`{e}aires Fossiles} (Eyrolles,
Paris, 1991).

\bibitem{Shl76}A. Shlyakhter,  Nature {\bf 264}, 340 (1976).

\bibitem{Fuj00}Y. Fujii et al.,  Nucl. Phys. {\bf B573}, 377 (2001), revisited in  
arXiv:hep-ph/0205206, 19 May 2002.

\bibitem{Dam96}T. Damour and F. Dyson,  Nucl. Phys. {\bf B480}, 37 (1996).

\bibitem{Chris06}C. R.  Gould, E. I.  Sharapov, 
and S. K. Lamoreaux,   Phys. Rev. C {\bf 74}, 024607 (2006).

\bibitem{Petr06}Yu. V. Petrov, A. I. Nazarov, M. S. Onegin, V. Yu. Petrov, 
and E. G. Sakhnovsky,   Phys. Rev. C {\bf 74}, 064610 (2006).

\bibitem{Oneg10}M. S. Onegin, 
 arXiv:1010.6299v1 [nucl-th], 29 Oct 2010.

\bibitem{Flam09}V. V. Flambaum, R. B. Wiringa, Phys. Rev. C {\bf 79}, 
034302 (2009).

\bibitem{Webb11}J. K. Webb et al., Phys. Rev. Lett. {\bf 107}, 191101 (2011).

\bibitem{Holli81}P. Holliger and  C. Devillers,  Earth and Planetary Science Letters, {\bf 42}, 76 (1981). 

\bibitem{Hida98}H. Hidaka and P. Holliger,   Geochimica et Cosmochimica Acta, {\bf 62}, No. 1, 891 (1998). 


\bibitem{West58}C. H. Westcott, Report CRRP-787,  Chalk River Laboratory, Canada, 1958.  

\bibitem{NNDC}Abundance and lifetime data from the National Nuclear Data Center, Brookhaven National Laboratory.

\bibitem{Mug06}  S.F.  Mughabghab, {\em Atlas of Neutron Resonances, Fifth Edition: Resonance 
Parameters and Thermal Neutron Cross Sections, Z=1 - 100}
(Elsevier, Amsterdam, 2006).

\bibitem{DeLa88} J. R.  De Laeter, B. J. Allen, G. C. Lowenthal, and J. W.  Boldeman, 
Astron.  Astrophys. {\bf 9}, 7  (1988). 

\bibitem{Hei08} M. Heil et al., Astrophys. J. {\bf673}, 434 (2008)
 
\bibitem{Wiss06}K. Wisshak, F. Voss,  F. K\"{a}ppeler, L. Kazakov,   
 Phys. Rev. C {\bf 73}, 015807 (2006).

\bibitem{Klay91}N.  Klay, F. K\"{a}ppeler, H. Beer et al.,  
Phys. Rev. C {\bf 44}, 2801 (1991).

\bibitem{Bast60} A. H.  Baston, J. C.  Lisle, G. S. G. Tuckey,
 J. Nucl. Energy Part I, {\bf 13}, 35 (1960).

\bibitem{Sears92} V. F. Sears, Neutron News, {\bf 3:3}, 26  (1992).

\bibitem{Dan11}M. J. Trbovich et al. Nucl. Sci. Eng. {\bf 161}, 303 (2009).

\bibitem{Gau96}F. Gauthier-Lafaye, P. Holliger, and P.-L. Blanc,  Geochimica et Cosmochimica Acta, {\bf 60}, No. 23, 4831 (1996). 

\bibitem{Thrane} K. Thrane, J. N. Connelly, M. Bizzarro, B. S. Meyer and 
L.S. The, Astrophys. J. {\bf 717}, 861 (2010).

\bibitem{Mohr09}P. Mohr, S. Bisterzo, R. Gallino,   F. K\"{a}ppeler, U. Kneissl, and   N. Winckler, Phys. Rev. C {\bf 79}, 045804 (2009).




\end{thebibliography}
\end{document}